\documentclass[10pt,conference]{IEEEtran}

\usepackage[utf8]{inputenc}   
\usepackage{cite}             
\usepackage{graphicx}
\usepackage{dcolumn}
\usepackage{bm}
\usepackage{hyperref}
\usepackage{amsmath,amssymb}  
\usepackage{braket}
\usepackage[capitalize]{cleveref}
\usepackage[caption=false]{subfig} 
\usepackage{color}
\usepackage{bm}
\newcommand{\refsec}[1]{Sec.~\ref{#1}}
\newcommand{\reffig}[1]{Fig.~\ref{#1}}
\newcommand{\refeq}[1]{Eq.~\ref{#1}}
\begin{document}

\title{Trainable Quantum Neural Network for Multiclass Image Classification with the Power of Pre-trained Tree Tensor Networks}
%
\author{
\IEEEauthorblockN{Keisuke Murota, Takumi Kobori}
\IEEEauthorblockA{Department of Physics, The University of Tokyo, Tokyo, Japan\\
keisuke.murota@phys.s.u-tokyo.ac.jp, takumi.kobori@phys.s.u-tokyo.ac.jp}
%
}
\maketitle

\begin{abstract}
Tree tensor networks (TTNs) offer powerful models for image classification.
While these TTN image classifiers already show excellent performance on classical hardware,
embedding them into quantum neural networks (QNNs) may further improve the performance by leveraging quantum resources.  
However, embedding TTN classifiers into QNNs for multiclass classification remains challenging.
Key obstacles are 
the high-order gate operations required for large bond dimensions and 
the mid-circuit postselection with exponentially low success rates necessary for the exact embedding. 
In this work, to address these challenges, we propose forest tensor network (FTN)-classifiers, 
which aggregate multiple small-bond-dimension TTNs.
This allows us to handle multiclass classification without requiring large gates in the embedded circuits.  
We then remove the overhead of mid-circuit postselection by extending the adiabatic encoding framework to our setting 
and smoothly encode the FTN-classifiers into a quantum forest tensor network (qFTN)-classifiers.
Numerical experiments on MNIST and CIFAR-10 demonstrate that we can successfully train FTN-classifiers and encode them into qFTN-classifiers,
while maintaining or even improving the performance of the pre-trained FTN-classifiers.
These results suggest that synergy between TTN classification models and QNNs can provide a robust and scalable framework for multiclass quantum-enhanced image classification.
\end{abstract}


\section{Introduction}
Tensor networks (TNs) provide a powerful framework for representing and manipulating high-dimensional data by decomposing large tensors into networks of low-rank factors. 
After being developed in the field of condensed matter physics~\cite{White1992,Verstraete2008,Orus2014,Orus2019,Cirac2021},
they have found diverse applications, 
including applied mathematics~\cite{Oseledets2011,khoromskij2012,cichocki2014, nunez2022}, 
quantum chemistry~\cite{chan2002, chan2011, baiardi2020, nakatani2018}, 
the simulation of quantum circuits~\cite{Pan2022, Liu2021, Patra2024, Fu2024, Vidal2003}, 
and machine learning (ML) tasks, such as generative modeling (unsupervised learning)~\cite{Han2018, Cheng2019, Miller2021},
and classification (supervised learning)~\cite{Stoudenmire2016, Evenbly2019, Reyes2020, Reyes_2021}. 
Recently, encouraged by the synergy between TNs and quantum circuits, and by the strong performance of TNs in ML tasks,
there is a growing trend toward embedding pre-trained TN-based models 
into quantum circuits~\cite{Rudolph2023,Iaconis2024,Wall2021-generative,Rudolph2024,Ran2020,Malz2024,Manabe2024,Sugawara2025,Green2025}. 
In this classical-quantum hybrid approach, 
classically trained TNs provide high-quality initializations for the quantum neural networks (QNNs)~\cite{Beer2020, Abbas2021, Kwak2021}.\
Once embedded, additional quantum gates are introduced, and the entire circuit is retrained on quantum hardware to move beyond the realm of classically simulatable quantum circuits. 
This allows us to take advantage of genuine quantum resources and potentially exceed the performance of purely classical ML models~\cite{Rudolph2023}. 
This synergetic strategy has already shown promise in contexts such as variational quantum eigensolvers (VQEs)~\cite{Peruzzo2014} and quantum generative modeling~\cite{Perdomo-Ortiz2018}.
Since the multiclass image classification is one of the most important and challenging tasks in ML,
an important next step is to focus on embedding TN-based multiclass image classifiers into QNNs.
In particular, we focus on tree tensor networks (TTNs) as our TN architecture~\cite{Shi2006}, given their proven success in capturing the key correlations within image data~\cite{Chen2023,Nie2025,Liu2019,Hikihara2023,Harada2025}.
In this way, we can expect to establish a blueprint for a new class of quantum-enhanced machine learning methods specifically designed for multiclass image classification.
\begin{figure*}
    \centering
    \includegraphics[width=\linewidth]{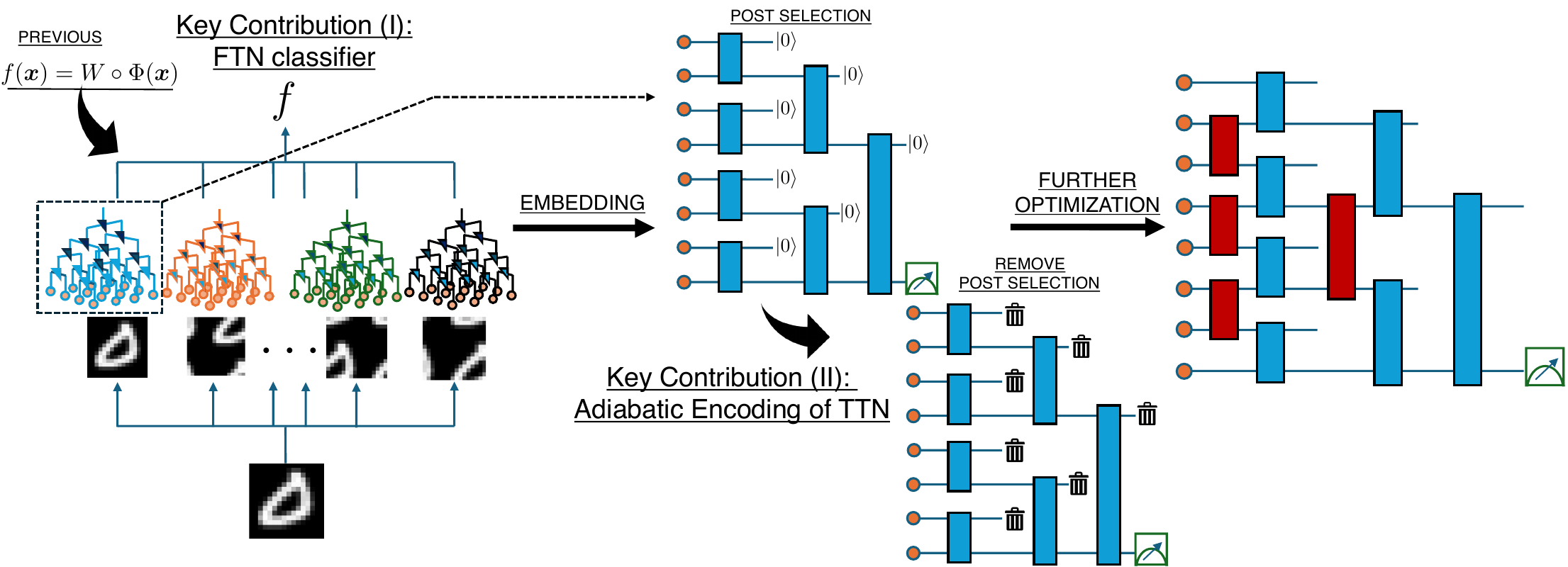}
    \caption{
    Overview of our proposed method and key contributions.
    In the context of the TN-QNN synergetic learning framework, the typical pipeline consists of three steps: 
    (1) training a classical TN-based ML model for initialization of QNN, 
    (2) embedding the pre-trained TN model into a quantum circuit, and 
    (3) further training the QNN model directly on quantum circuits.
    Our contributions address the first two steps: 
    (I) we propose FTN-classifiers, classical TTN-based learning models for multiclass image classification 
    that are efficient and easy to embed into quantum circuits, and 
    (II) we extend the adiabatic encoding framework to TTNs to remove the need for postselection. 
    The last step should be conducted in future work.
    In this paper, we refer to the quantum circuits that have the same underlying structures as TTN 
    as qTTN-circuit, and the corresponding classifier as qTTN-classifier.
    The same applies to the notation of qFTN-circuit and qFTN-classifier.
    }
    \label{fig:overview}
\end{figure*}

Despite the promising prospects, 
the direct embedding of TTN-based classifiers (TTN-classifiers) into QNNs for multiclass datasets remains largely unexplored 
and have several challenges:
\begin{itemize}
    \item \textbf{Binary-to-multiclass extension:} 
    While TTN-classifiers for classical hardware can be extended to multiclass tasks by simply increasing the bond dimension,
    attempting to compile such large bond dimension TTNs into quantum circuits requires high-order quantum gates.
    Such large gate operations are not preferred to be implemented in current quantum hardware, considering that large unitary gates need
    to be decomposed into a sequence of smaller gates~\cite{Vartiainen2004, Krol2024, Nakanishi2024, Cross2019}.
    Although some studies have discussed how to use TTN-based QNNs for multiclass image classification~\cite{wall2021,lazzarin2022,araz2022}, 
    there is as yet no established strategy, and the problem continues to be non-trivial.
    \item \textbf{Mid-circuit postselection:} 
    Although a classically pre-trained TTN-classifier can be exactly embedded into a quantum TTN-circuit (qTTN-circuit),
    a naive embedding requires
    postselection steps during the circuit execution~\cite{huggins2019,rieser2023,kodama2025}.
    The success probability of this postselection 
    can decrease exponentially with system size, making straightforward 
    implementations on quantum hardware impractical.
\end{itemize}
In this work, we propose a new method to encode pre-trained TTN-based models for multiclass image classification into QNNs 
by addressing the two challenges mentioned above.
Our main contributions are as follows:
\begin{itemize}
    \item \textbf{Handling multiclass image classification:}
    Instead of using a single TTN-classifier with large bond dimensions,
    we introduce forest tensor network classifiers (FTN-classifiers), 
    which use multiple TTN-classifiers with different pixel connectivities and small bond dimension,
    and then aggregate the outputs of each TTN to predict the multiclass label.
    Importantly, if we allow postselection, we can rigorously embed FTN-classifiers into quantum circuits.
    \item \textbf{Removing mid-circuit postselection:}
    To eliminate the overhead associated with mid-circuit postselection, we extend the 
    adiabatic encoding framework from~\cite{keisuke2025}—originally designed for 
    binary-label matrix product states (MPS)-classifiers—to the more general TTN setting. 
    This enables a 
    systematic embedding of pre-trained TTN-classifiers into postselection-free 
    QNN architectures.

    \item \textbf{Numerical results:}
    We demonstrate the trainability and expressiveness of our proposed method on the MNIST 
    and CIFAR-10 datasets. 
    Our findings show that we can successfully pre-train FTN-classifiers
    and convert them into postselection-free qFTN-classifiers.
    Notably, we confirmed that the qFTN-classifiers achieve over 99\% training accuracy on MNIST 
    and over 96\% training accuracy on CIFAR-10.

\end{itemize}
The graphical overview of our proposed method and key contributions is shown in~\cref{fig:overview}.

\section{Multiclass Classification with TTN-Classifiers}\label{sec:multiclass}
This section first reviews conventional TN-based classification methods and then introduces our proposed TTN-based classifiers for multiclass image classification.
Among the various approaches of TN methods for ML, we focus on the non-linear kernel learning utilizing TN. 
In the simple case, the prediction function $f(\bm{x})$ for $d$-class classification with an $n$-dimensional input data $\bm{x}\in \mathbb{R}^n$ can be expressed as
\begin{equation}
    f(\bm{x}) = W \circ \Phi(\bm{x})
\end{equation}
where $\Phi: \mathbb{R}^n\rightarrow \mathcal{S}$ is the non-linear feature map that embeds the input into a high-dimensional feature space $\mathcal{S}$ and $W: \mathcal{S}\rightarrow\mathbb{R}^d$ is a trainable linear model. 
The predicted label is chosen as the index corresponding to the maximum value of $f$.
In the context of image classification with TNs, we use exponentially-high-dimensional feature space $\mathcal{S}=\mathbb{R}^{2^n}$,
and a feature map can be defined as a tensor product of a two-dimensional local feature map for each input element. In this work, we employ the following normalized local feature map $\phi: \mathbb{R}\to\mathbb{R}^2$:
\begin{equation}\label{eq:feature_map_local}
    \phi(x) = \frac{1}{\sqrt{x^2 + (1-x)^2}}\begin{pmatrix}
        x \\ 1-x
    \end{pmatrix}
\end{equation}
which ensures unit norm for each mapped vector when $x\in[0,1]$.
We can use other feature maps such as $\phi(x) = {(\cos(\tfrac{\pi}{2}x), \sin(\tfrac{\pi}{2}x))}^T$ 
as long as they are normalized,
but the rest of the discussion is the same.
The normalization in $\phi$ is required for embedding it into quantum circuits.
The global feature map $\Phi: \mathbb{R}^n\rightarrow \mathbb{R}^{2^n}$ is defined as the tensor product over all components:
\begin{equation}\label{eq:feature_map}
    \Phi(\bm{x}) = \bigotimes_{i}^{n}\phi(x_i).
\end{equation}
For an input image data of size $L_h\times L_w$, the input dimension becomes $n=kL_hL_w$, where $k$ denotes the number of channels per pixel, for example $k=1$ for grayscale images and $k=3$ for RGB images.
Embedding input data $\bm{x}$ into a high-dimensional space through this feature map 
allows for the construction of separating hyperplanes that can discriminate complex patterns.
However, computing the prediction function directly is infeasible 
as it requires an exponentially large vector-matrix product.
By leveraging the efficient data compression capabilities of TNs, 
we can approximate this computation with high accuracy.
\subsection{Tensor Network}
TNs are computational frameworks that efficiently represent and manipulate high-dimensional data by decomposing large tensors into networks of smaller, interconnected multi-dimensional tensors.
Each node in a tensor network represents a tensor, and each tensor has edges, called legs, corresponding to its number of dimensions.
The interconnections between tensors represent operations similar to matrix multiplications, known as tensor contractions.
For example, when two 3-leg tensors $A_{ijk}$ and $B_{klm}$ are connected through a shared leg indexed as $k$, 
which has dimension $n$ referred to as bond dimension, a contraction is performed over that index as follows:
\begin{equation}
    C_{ijlm} = \sum_{k=0}^{n-1}A_{ijk}B_{klm}
\end{equation}
resulting in a new 4-leg tensor $C_{ijlm}$.
Without any restriction on the bond dimensions, tensor networks can represent the original tensor exactly.
However, achieving such an exact representation typically demands exponentially large bond dimensions.
By limiting each bond dimension to a fixed size, say $\chi$, through sophisticated TN techniques and under specific network structures, 
one can achieve efficient data compression.
\subsection{Basic tree tensor network classifier for image data}\label{sec:basic_TTN}
Within the TN framework, the tensor product in \refeq{eq:feature_map} can be represented 
exactly by a tensor network, since each two-dimensional local feature map corresponds 
to a one-leg tensor with bond dimension two. The remaining task is to efficiently 
represent the exponentially large linear map $W$ via a suitable TN decomposition. 
Depending on the spatial structure of the target system, there exist various TN architectures for representing $W$,
including MPS~\cite{Orus2014}, TTNs~\cite{Shi2006}, multi-scale entanglement renormalization ansatz (MERA)~\cite{Vidal2008}, and projected entangled pair 
states (PEPS)~\cite{Verstraete2004}. 
Because the choice of TN architecture heavily influences learning performance, it is crucial to select the architecture carefully.

\begin{figure}
    \centering
    \includegraphics[width=\linewidth]{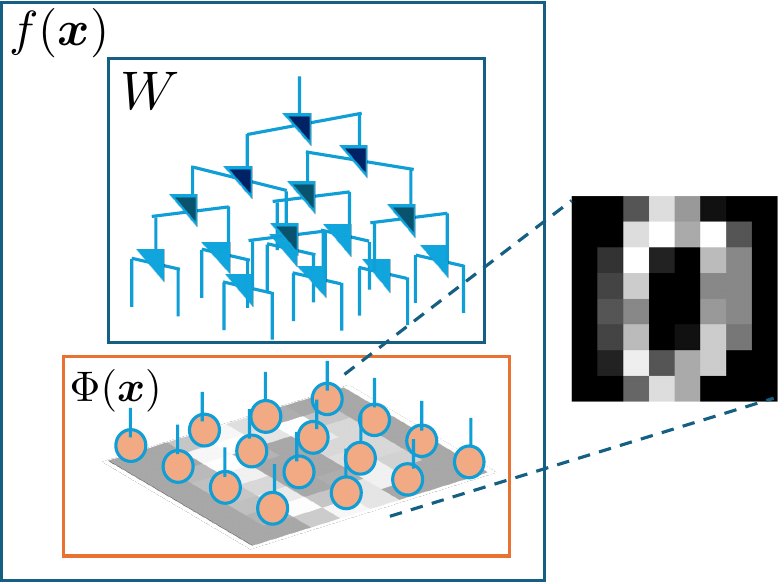}
    \caption{Conceptual diagram of basic TTN-classifier.}
    \label{fig:ttn_classifier}
\end{figure}

In this work, we employ the TTN architecture with the underlying tree data structure 
being a height-balanced binary tree whose internal nodes each have exactly two children~\cite{Cheng2019}. 
This architecture leverages the hierarchical structure and computational efficiency of TTNs, 
making it well-suited for capturing the complex 2D spatial correlations in image data~\cite{Harada2025}.
A conceptual diagram of the resulting TTN-based prediction function $f(\bm{x})$ can be found in 
\reffig{fig:ttn_classifier}.
To fully utilize the computational resources of TTNs, we assume the input images are squared $L = L_h = L_w$ and 
$L$ is a power of two.
In the TTN-classifier, the bond dimensions are determined as follows:
For the input feature map, the physical bonds associated with each pixel are grouped to give a bond dimension of $2^k$.
In the TTN structure, all intermediate three-leg tensors (except for the top tensor) are typically set to have dimensions $\chi\times\chi\times\chi$, with $\chi$ being a fixed bond dimension.
The top tensor, representing the final output for $d$-class classification, has dimensions $d\times\chi\times\chi$. 
When we look at the computational complexity, the space complexity for calculating $f(\bm{x})$ is $O(N\chi^3)$, 
where $N = L^2$ is the number of pixels, or equivalently, the dimension of the input data.
The time complexity is also $O(N\chi^3)$ for single-core computation, and it can be reduced to $O(\chi^3\log N)$ with parallelism.
As a result, the TTN-classifier enables efficient computation both in terms of memory usage and computation time and 
it is well-suited for large-scale image classification tasks.

\subsection{Forest tensor network classifier for multiclass classification}
Although the basic classifier in Sec.~\ref{sec:basic_TTN} can be employed for multiclass classification, several technical challenges are anticipated when embedding it into quantum circuits. 

The first challenge is the limitation on the bond dimension $\chi$, to prevent unitary gates from becoming excessively large.
As will be explained in Sec.~\ref{sec:exact-representation}, embedding a $\chi\times\chi\times\chi$ tensor into a quantum circuit generally requires a unitary gate acting across $2\log\chi$ qubits.
Therefore, if $\chi$ becomes too large, the corresponding unitary gates after embedding also increase in size.
The decomposition and synthesis of such large unitary gates are among the most actively researched topics in quantum compilation, 
but their impact on overall circuit performance remains uncertain~\cite{Iten2016, Krol2024, Shende2004}.
Therefore, we need to avoid using a large bond dimension $\chi$ 
if we aim to embed the TTN-classifiers into quantum circuits as QNN initializations.


The second challenge happens when the number of classification classes $d$ is not a power of two, like in the case for MNIST or CIFAR-10 datasets.
In this case, embedding the top tensor into a quantum circuit requires padding the data appropriately such that the bond dimension becomes a power of two, which is necessary for exactly representing the top tensor as a unitary operator.
However, this padding requires discarding certain measurement outcomes, which increases the number of samples required for reliable classification by a constant factor.

The third challenge is the potential reduction in the expressive power of the TTNs due to the restriction on $\chi$.
As mentioned earlier, it is preferable to use a model with a small $\chi$ for the sake of quantum circuit compatibility.
In the TN framework, the bond dimension $\chi$ controls how much correlation between pixels can be captured and compressed.
A small $\chi$ may lead to over-compression, limiting the model’s expressive power.

\begin{figure}
    \centering
    \includegraphics[width=\linewidth]{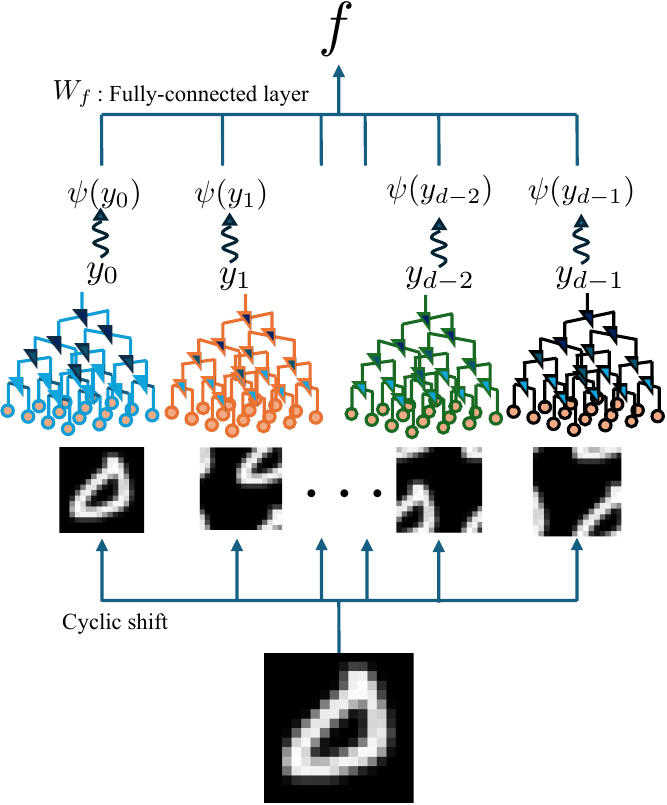}
    \caption{Schematic diagram of our proposed FTN-classifier.
    We use multiple TTN-classifiers to extract diverse features from a single image.
    }
    \label{fig:ttn_multi}
\end{figure}

To overcome these challenges, we introduce a novel TTN-based classification model called forest tensor network classifier~(FTN-classifier), as shown in \reffig{fig:ttn_multi}.
Our FTN-classifier consists of four layers: the first is a cyclic shift layer; the second performs feature extraction using TTNs; the third applies a non-linear mapping that mimics quantum circuit measurement; and the fourth is a fully connected linear layer.
The basic TTN-classifier in Sec.~\ref{sec:basic_TTN} serves not only to compute the prediction function but also to extract meaningful features from the original image data.
In addition, we restrict the bond dimension $\chi$ of each TTN-classifier to $2^k$, which is just $2$ for grayscale images and $8$ for RGB images,
so that the operation on quantum hardware is feasible after embedding.
However, such restrictions reduce the amount of information extracted from the input image, which in turn leads to a degradation in prediction performance.
To address this limitation, we introduce multiple independent TTNs as feature extractors in our proposed model.
While the number of TTNs is a tunable hyperparameter, in this work, we fix it to $d$, the number of classes.
In the cyclic shift layer, each TTN-classifier applies the horizontal or vertical shift to the input images before processing them as input quantum states.  
This operation effectively endows each TTN classifier with a slightly different connectivity pattern. 
For example, in~\cref{fig:ttn_classifier}, one of the three-leg tensors at the bottom layer 
originally connects the pixels of the coordinates (0,0) and (1,0). 
However, by shifting the image, e.g.\ one pixel vertically, the pixels at (0,0) and (3,0) become directly connected instead, which effectively corresponds to modifying the pixel connectivity within the TTN.\
Hence, with the cyclic shift, we can efficiently extract diverse features from a single image without modifying the TTN architecture.
Note that each TTN has independent parameters.

After extracting features by the basic TTN-classifier, we apply a non-linear map $\psi:\mathbb{R}^{2^k}\to\mathbb{R}^{2^k}$, which mimics quantum circuit measurement: 
\begin{equation}
    \psi(\bm{y}) = (y_0^2, y_1^2,\cdots, y_{2^k-1}^2) / \sum_i y_i^2.
\end{equation}
This is necessary to rigorously reproduce the output of the basic TTN-classifier using quantum hardware.
Finally, after aggregating the outputs of all $d$ TTNs, a linear layer $W:\mathbb{R}^{d2^k}\to\mathbb{R}^{d}$ follows them, which performs classification on $d$ classes.
The total number of parameters in the model is $(N-1)2^{3k}d+2^kd^2$ for 
$d$-class classification on an image of size $N = L^2$ pixels with $k$ channels.
This value remains at most several tens of times the number of pixels when $d$ and $k$ are small, and thus the model can be considered sufficiently small.
With this model and the embedding method for quantum circuits described later, efficient quantum circuit embedding can be achieved while also maintaining high classification performance, as demonstrated in \refsec{sec:numerical-results}.
\section{Exact Representation of TTN-Classifier as a QNN with postselection}\label{sec:exact-representation}

Because an FTN-classifier is composed of multiple TTN-classifiers, 
its quantum counterpart (qFTN-classifier) simply consists of multiple qTTN-classifier 
modules arranged according to the same ``forest'' structure, followed by a classical 
linear layer.
To exactly reproduce the output of FTN-classifiers using qFTN-classifiers (with postselection), 
we just need to embed each TTN-classifier as a qTTN-classifier with the following three steps.

\textit{Step 1.} We first perform the canonical decomposition of each TTN-classifier. 
Since TTNs generally have residual gauge degrees of freedom on their internal bonds, we can suitably adjust these degrees of freedom so that all of the tensors, except for the top tensors, become isometries. 
Note that this adjustment does not affect the overall output of the TTN-classifiers, 
and FTN-classifiers.
\begin{figure}
    \centering
    \includegraphics[width=0.2\textwidth]{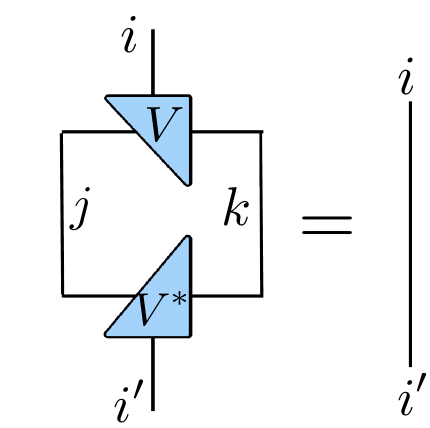}
    \caption{TN diagram of isometry condition for three-leg tensor $V_{ijk}$. In TN diagram notation, the identity tensor is represented as a simple line.}\label{fig:isometry}
\end{figure}
A TTN in canonical form consists of a single canonical center, which is a diagonal matrix, and a set of isometries.
As an example of the isometry condition for a matrix $V_{i, (j,k)}$, after grouping the last two legs of three-leg tensor $V_{ijk}$, the condition is given by
\begin{equation} 
\sum_{j,k} V_{i', (j,k)}^* V_{i, (j,k)} = \delta_{i,i'}.
\end{equation}
where $V^{*}$ is the complex conjugate of $V$.
This schematic illustration is shown in \reffig{fig:isometry}. 
This condition can be regarded as a generalization of the unitary condition.
General TTNs can be transformed into their canonical form using singular value decomposition (SVD).
In the case of a full binary TTN, this transformation is achieved by successively applying isometric transformations via SVD from the bottom layer to the top.
Specifically, each three-leg tensor $A_{ijk}$ of the bottom layer is reshaped into matrix $A_{i(jk)}$ of shape $\chi\times\chi^2$, and then decomposed as 
\begin{equation}   A_{i(jk)}=\sum_{ll'}U_{il}\Lambda_{ll'}V_{l'(jk)}
\end{equation}
by SVD.
By replacing $A_{ijk}$ with $V_{l'jk}$, the resulting tensor becomes isometry. 
The remaining part, $U_{il}\Lambda_{ll'}$, is then contracted with the tensor at the next higher layer.
By repeating this isometrization process from the bottom to the top, all tensors in the TTN, except for the top tensor, are converted into isometries.
For the top tensor, the canonical form can also be obtained via SVD; however, in the context of quantum embedding, we intentionally leave the top tensor non-isometric during the preprocessing step.

\textit{Step 2.} We then embed each isometry into a unitary operator.
Concretely, we have an isometry $V$ of dimension $\chi \times \chi^2$, with $\chi = 2^k$.
In this case, we can embed $V$ of the size $\chi^2 \times \chi$ into a full unitary $U$ of the size $2\chi^2 \times 2\chi^2$ via
\begin{equation}
    U = \begin{pmatrix}
        V \\
        V^\perp
    \end{pmatrix},
    \label{eq:embedding_isometry}
\end{equation}
where the rows of $V^\perp$ form an orthonormal basis orthogonal to those of $V$ and has dimension $\chi^2 \times (\chi^2 - \chi)$.
\begin{figure}
    \centering
    \includegraphics[width=0.29\textwidth]{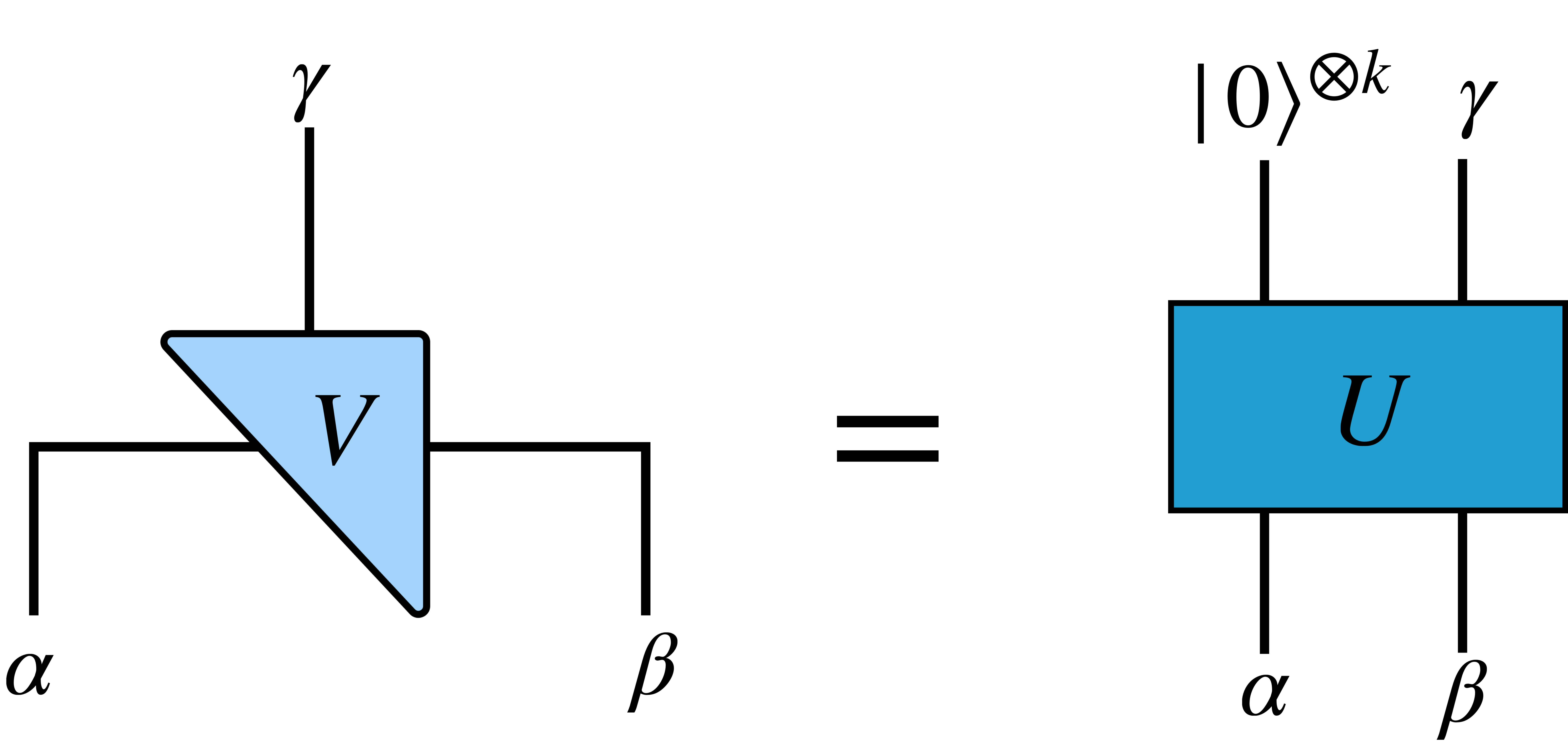}
    \caption{Illustration of the embedding procedure. $V$ is an isometry of dimension $N \times M$, and $U$ is a unitary of dimension $N \times N$.
    In order to exactly encode $V$ into $U$, we need to perform postselection on the first $k$ qubits after applying $U$.
    $\alpha, \beta, \gamma$ correspond to the $k$ qubit states.}\label{fig:embedding}
\end{figure}
This correspondence is illustrated in~\cref{fig:embedding}.
Because of the isometry condition, the given unitary matrix satisfies the unitary condition as expected.

To operate $V$ on an $\chi^2$-dimensional quantum state $\ket{\psi}$, we apply the unitary $U$ 
and then measure the first $k$ qubits in the computational basis, postselecting on the outcome 
$\ket{0} \cdots \ket{0}$. In that event, the normalized post-measurement state is
\begin{equation}
    \frac{V \ket{\psi}}{\|V \ket{\psi}\|}.
    \label{eq:post}
\end{equation}
By repeating this procedure for each isometry in the TTN-classifiers that is already in canonical form, 
we can realize the entire TTN-classifiers up to the second to the last layer
as a sequence of quantum gates with intermediate postselection. 
We still leave the top tensors not embedded into quantum gates, since they are not isometries.

\textit{Step 3.}
To embed the top non-isometric tensor $M$ into a unitary operator, we adopt a block-encoding approach~\cite{Low2019hamiltonian, chakraborty2019}.
First, we reshape the $\chi \times \chi \times \chi$ tensor $M$ into a $\chi^2 \times \chi$ matrix, 
followed by a proper rescaling to obtain $\tilde{M}$ to ensure that $\tilde{M}$.
Note that the rescaling of $M$ does not change the output of the multiclass TTN-classifier.
We embed $\tilde{M}$ into a $2\chi^2 \times 2\chi^2$ unitary matrix $U_b$ of the form
\begin{equation}
    U_b =
    \begin{pmatrix}
        \tilde{M} & C \\
        B & D
    \end{pmatrix},
    \label{eq:unitary_matrix}
\end{equation}
where the block $B$ of the size $\chi^2 \times \chi$ must satisfy
\begin{equation}
    B^\dagger B 
    =
    I_{\chi}
    -
    \tilde{M}^\dagger \tilde{M}.
    \label{eq:block_B_condition}
\end{equation}

\begin{figure}
    \centering
    \includegraphics[width=0.3\textwidth]{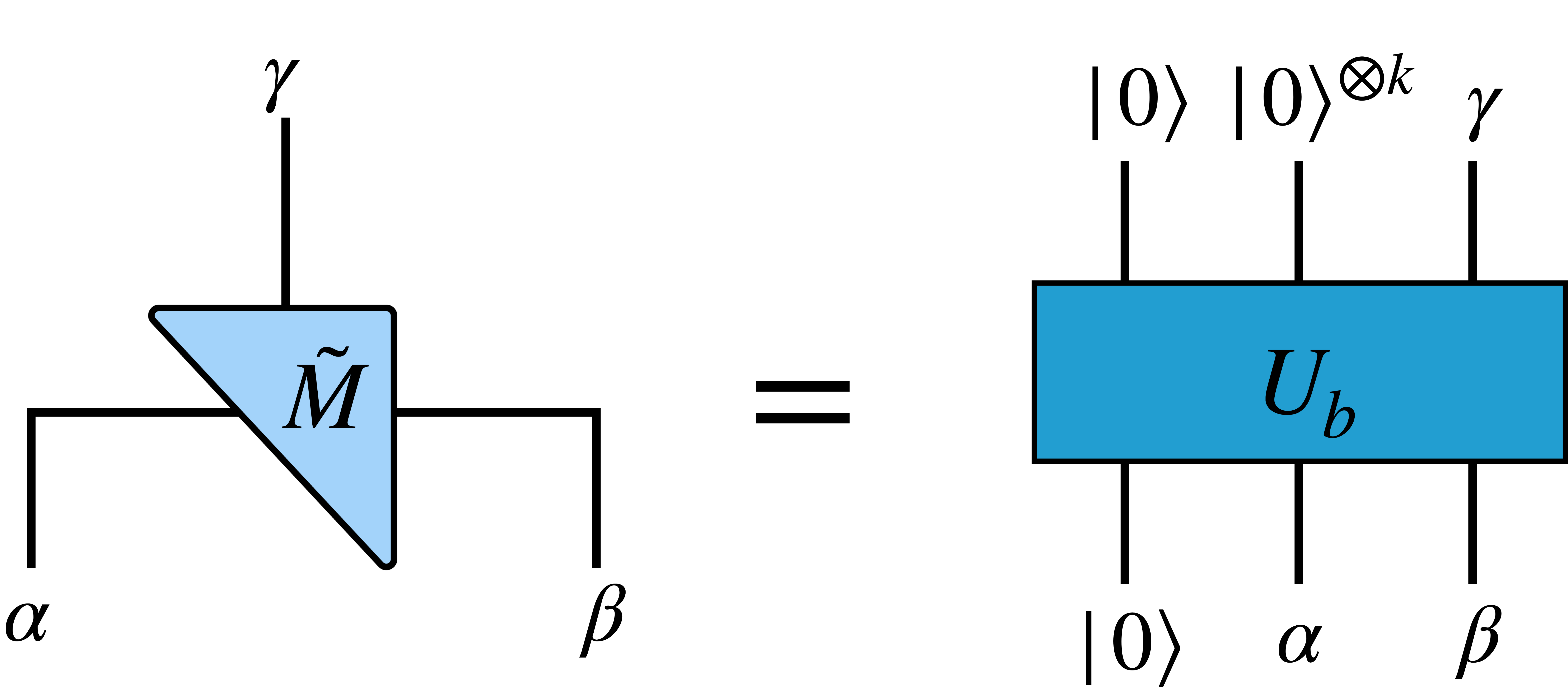}
    \caption{Illustration of the block encoding of $\tilde{M}$ of dimension $\chi \times \chi \times \chi$ into $U_b$ of dimension $(2\chi^2) \times (2\chi^2)$.
    The left line in $U_b$ corresponds to the ancilla qubit. The bottom left $\ket{0}$ represents the $\ket{0}$ initialization of the ancilla qubit.
    In addition to the postselection in $k$ qubits, we also postselect on the ancilla qubit.
    }\label{fig:block_encoding}
\end{figure}

From the fact that $B^\dagger B$ is always a positive-definite matrix, 
we deduce that the singular values of $\tilde{M}$ must be no greater than 1. 
Conversely, one can rescale $\tilde{M}$ so that its singular values do not exceed 1.
Next, diagonalize $B^\dagger B$ as
\begin{equation}
    B^\dagger B \;=\; X\,\Lambda\,X^\dagger.
    \label{eq:diagonalization}
\end{equation}
From this expression, we can construct $B$ as
\begin{equation}
    B 
    = 
    \sqrt{\Lambda}X^\dagger.
    \label{eq:rescaled_B}
\end{equation}
With this $B$, define
\begin{equation}
    W 
    \;=\;
    \begin{pmatrix}
        \tilde{M}\\[6pt]
        B
    \end{pmatrix},
    \label{eq:W}
\end{equation}
where $W$ is a $(2\chi^2)\times \chi$ isometry. 
To complete it into a full $(2\chi^2)\times(2\chi^2)$ unitary, 
we merely add the remaining $(2\chi^2 - \chi)$ columns orthogonal to $W$. 
A practical way to achieve this is via the QR decomposition applied to $W$.
To completely recover the operation of $\tilde{M}$ from the operation of $U_b$,
we need to prepare the ancilla qubit in the $\ket{0}$ state and perform postselection on $k+1$ qubits so that
all the qubits are in the $\ket{0}^{\otimes k + 1}$ state, as illustrated in~\cref{fig:block_encoding}.

By applying the above procedure, we can exactly embed TTN-classifiers into quantum circuits
using qTTN-circuits.
However, since each 3-leg tensor embedding involves postselection on several qubits, 
the overall success probability of the postselection-based qFTN-classifier may decrease exponentially with system size. 
This fundamental limitation motivates us to explore how to remove the 
postselection requirement, which we discuss in the next section.

\section{Removing postselection using Adiabatic Encoding}\label{sec:adiabatic-encoding}
As shown in \cref{sec:exact-representation}, directly embedding TTN-classifiers into quantum circuits 
requires exponentially costly postselection.
This issue is not unique to TTNs but arises for TN-based classifiers in general.
To circumvent this problem, previous research~\cite{keisuke2025} proposed an adiabatic encoding framework to encode MPS-classifiers into postselection-free QNNs,
without any loss in performance. 
In this work, we extend that framework to TTN-classifiers.

\begin{figure}
    \centering
    \includegraphics[width=0.5\textwidth]{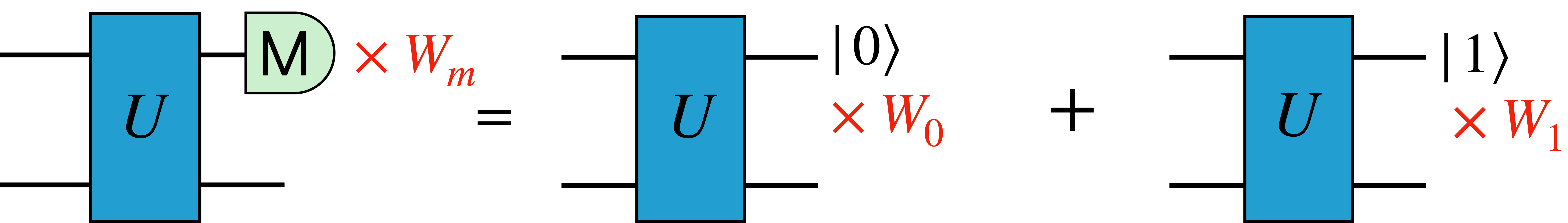}
    \caption{Illustration of the weighted quantum state.
    If the measurement outcome is $0$ or $\ket{0}$, we multiply the 
    final outcome by $W_0$ and if the outcome is $1$ or $\ket{1}$, we multiply the 
    final outcome by $W_1$.
    We confirm that at the end of the training, 
    the training loss and accuracy become $0.0176$ and $99.06\%$ respectively.
    }\label{fig:weighted_quantum_state}
\end{figure}
We begin by briefly reviewing the adiabatic encoding framework. 
As a preliminary, we introduce the concept of \emph{weighted quantum states}~\cite{Holmes2023}.
Rather than simply discarding outcomes or performing postselection, 
we assign a classical weight to each mid-circuit measurement outcome and the final outcome
is given by weighting the classical weights to the output quantum state.
Concretely, consider a general quantum instrument $\mathcal{E}_m$ that includes a measurement $M$ acting on an input density matrix $\rho_{\text{in}}$, 
where outcomes are labeled by $m$. 
We then define the weighted quantum state $\tilde{\tau}$ as
\begin{equation}\label{eq:weighted_quantum_state}
    \tilde{\tau} = \sum_{m} W_m \, p_m \, \frac{\mathcal{E}_m\bigl(\rho_{\text{in}}\bigr)}{\mathrm{Tr}\Bigl[\mathcal{E}_m\bigl(\rho_{\text{in}}\bigr)\Bigr]} 
    = \sum_{m} W_m \, \mathcal{E}_m\bigl(\rho_{\text{in}}\bigr),
\end{equation}
where $p_m = \mathrm{Tr}\bigl[\mathcal{E}_m(\rho_{\text{in}})\bigr]$ is the probability of obtaining outcome $m$, 
and $W_m$ is a user-defined weight associated with that outcome. 
~\cref{fig:weighted_quantum_state} illustrates this concept.
Since the scale of the quantum state does not matter in our case, we can set $W_m \leq 1$ for all $m$.
This rescaling also provides an alternative operational interpretation of the weighted 
quantum state in \cref{eq:weighted_quantum_state}. 
Specifically, one can realize $\tau$ by performing a mid-circuit measurement 
and then postselecting the measurement outcome $m$ with probability $W_m$. 
In other words, rather than deterministically accepting or rejecting each measurement outcome, 
one randomly decides to continue the quantum circuit run with a probability.
The success probability of this operation, or $1 - \text{rejection probability}$, is given by
\begin{equation}
    \label{eq:success_probability}
    P_{\text{success}} = \mathrm{Tr}\bigl[\tilde{\tau}\bigr] = \sum_{m} W_m p_m.
\end{equation}
You can immediately understand that $P_{\text{success}}$ reduces to the success probability of the standard postselection when $W_0 = 1$ and $W_1 = 0$,
and reduces to $1$ when $W_0 = 1$ and $W_1 = 1$,
which can be lead from the fact that $\sum_{m} \mathrm{Tr}\bigl[\mathcal{E}_m(\rho)\bigr] = \mathrm{Tr}\bigl[\rho\bigr] = 1$.

With this expression, the standard postselection-based operation and the discarding measurement operation (partial trace) 
can be regarded as two extreme cases of the weighted quantum states.
In one extreme, one only accepts outcomes when $\ket{0}$ is measured ($W_0 = 1$ and $W_1 = 0$),
while rejecting all others (the usual notion of postselection).
In the other extreme, one always accepts the measurement outcome ($W_m = 1$ for both $m=0$ and $m=1$), 
effectively discarding no measurement record (partial trace). 

\begin{figure}
    \centering
    \includegraphics[width=0.49\textwidth]{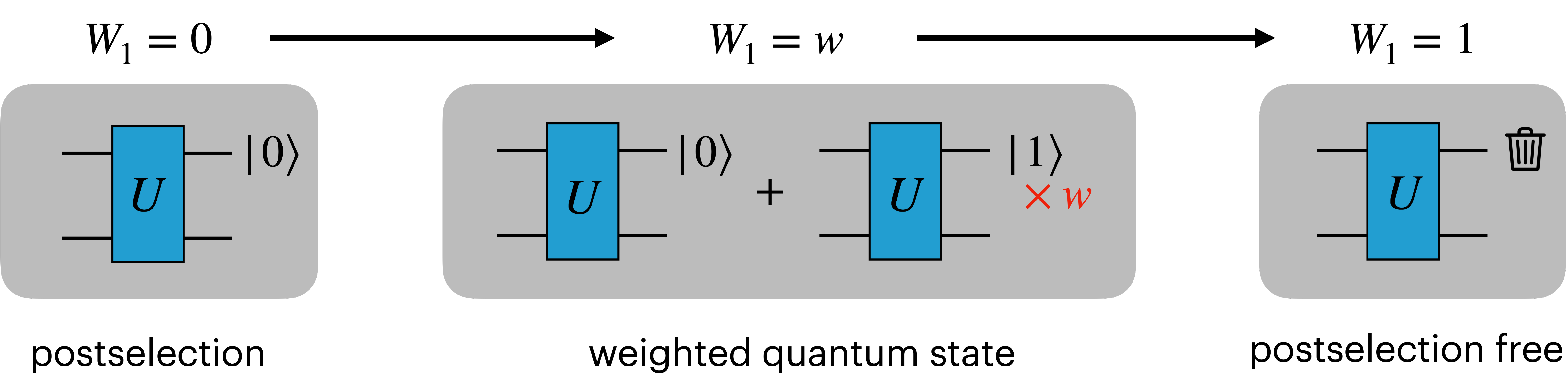}
    \caption{
        Illustration of gradual changes from the postselection-based circuit to
        the postselection-free circuit through the weighted quantum state.
        The trash bin in the right figure represents the discarded measurement outcomes (or partial trace).
    }\label{fig:adiabatic_encoding_wq}
\end{figure}
In the adiabatic encoding framework, we introduce a continuous weight parameter $W_1 = w \in [0,1]$,
that represents the probability of accepting the $\ket{1}$ outcome in the mid-circuit measurement.
We begin with $w=0$, where the circuits are equivalent to the postselection-based qTTN-classifiers,
which can be obtained by exactly embedding the TTN-classifiers into quantum circuits, as illustrated in~\cref{fig:adiabatic_encoding_wq}. 
We then gradually increase $w$ from 0 to 1 in small increments $\Delta w$.
At each $w$, we optimize the parameters of the QNNs  
until the loss function falls below a predetermined threshold. 
Then we increment $w$ by $\Delta w$, choosing $\Delta w$ in such a way that the loss does not 
increase excessively after proceeding to $w + \Delta w$.
By the time $w$ reaches 1, the circuit no longer requires any postselection.
After making all the qTTN-classifiers postselection-free, we can successfully construct
high-quality qFTN-classifiers that are fully implementable on quantum hardware.

Finally, let us remark on the computational resources required to 
calculate the output for a single qTTN-classifier on classical computers
and quantum computers. 

\textbf{Classical computer.}
Unlike the classical TTN-classifier from \cref{sec:multiclass}, 
each leg of a qTTN carries a $\chi \times \chi$ density matrix as input, 
so the qTTN can effectively be viewed as a TTN with bond dimension $\chi^2$. 
Consequently, the simulation cost scales as $O\bigl(N(\chi^2)^3 \bigr) = O(N\chi^6 )$, 
which is significantly larger than the $O(N\chi^3)$ complexity for TTN-classifiers.
If we can employ parallelization, 
this can be reduced to $O(\chi^3 \log N)$.
\textbf{Quantum computer.}
A direct implementation of a qTTN-classifier on quantum hardware 
can use $O(N)$ qubits with a circuit depth of $O(\log N)$. 
Alternatively, if one aims to minimize the number of qubits, 
the circuit allows qubit-efficient form which only requires $O(\log N)$ qubits at the cost of increasing the circuit depth to $O(N)$,
by leveraging mid-circuit measurements and resets~\cite{rieser2023}.
Hence, for a $d$-class classification task, the qFTN-classifier can be instantiated 
using $d$ separate $O(\log N)$-qubit circuits on a quantum hardware,
which demonstrates the practicality of our approach in the near-term regime.

\section{Numerical Results}\label{sec:numerical-results}
We tested our method on two standard datasets: MNIST and CIFAR-10. 
MNIST is a well-known dataset of handwritten digits (0-9) of size $28\times28$ that consists of 60000 training and 10000 test samples.
On the other hand, CIFAR-10 is a collection of RGB color images of size $32\times32$ that consists of 50000 training and 10000 test samples.
To calculate the loss, we first apply a softmax function to the outputs of the final fully connected layer to derive class probabilities,
followed by computing the negative log-likelihood on these probabilities.
We first trained an FTN-classifier using the Adam optimizer, then exactly embedded the trained FTN-classifier into a quantum circuit 
using the exact procedure described in \cref{sec:exact-representation}. 
Although this embedded quantum circuit requires postselection, 
we gradually increased $w$ by following the framework introduced 
in \cref{sec:adiabatic-encoding}, so that by the time $w=1$, 
postselection is no longer necessary. 
Since we need to further optimize the unitary gates in the QNNs during adiabatic encoding,
we use Riemannian stochastic gradient descent~\cite{Bonnabel2013} implemented by 
the \texttt{geooptp} package~\cite{geoopt2020kochurov} in Python.

\subsection{Results on MNIST}
First, we present the results of applying our method to the MNIST dataset. 
Although MNIST consists of $28 \times 28$ grayscale images in our setting, 
we downscale each image to $16 \times 16$ and rescale the pixel values to $[0,1]$. 
We then embed them into single-qubit quantum states using the feature map in \cref{eq:feature_map_local}.

\begin{figure}[htp]
    \centering
    \includegraphics[width=0.5\textwidth]{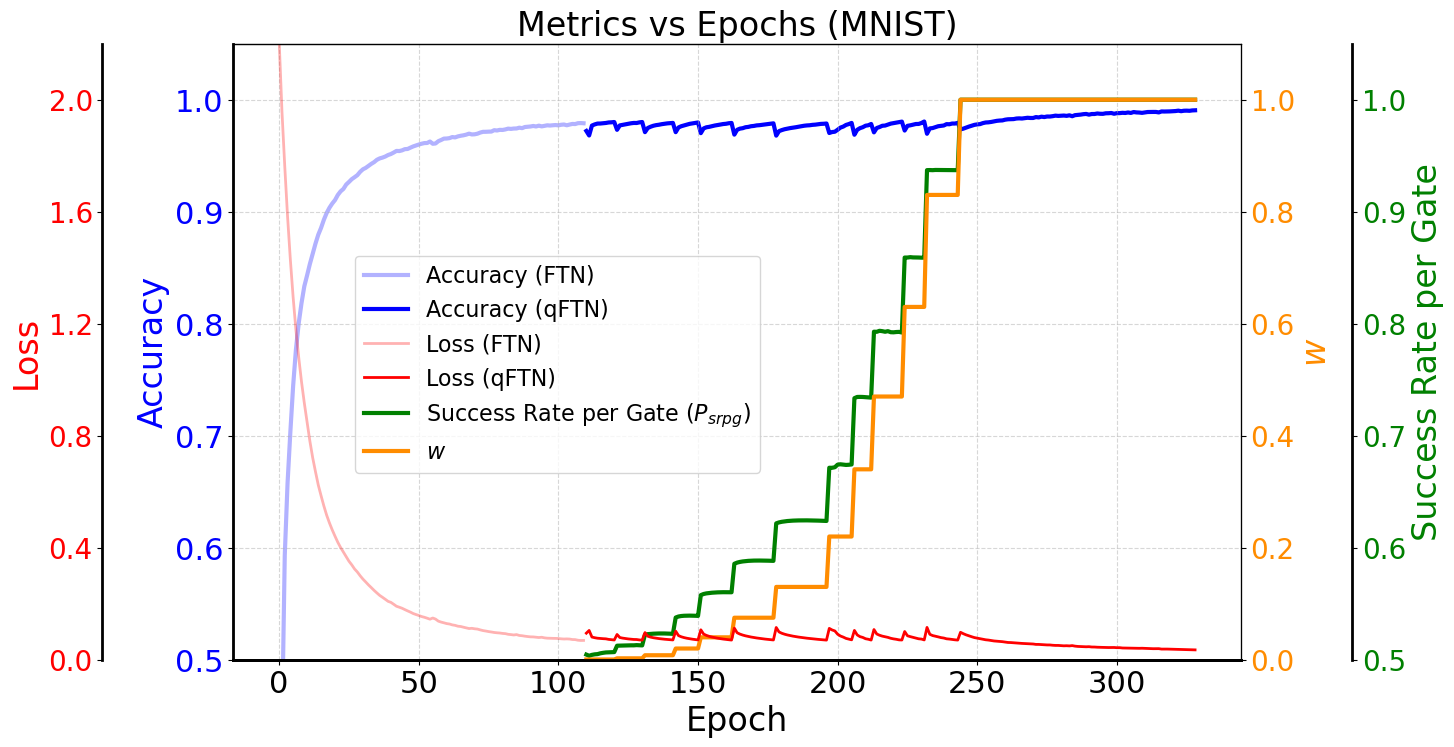}
    \caption{
        Training and embedding results for the MNIST dataset. 
        The lighter-colored curves correspond to training the FTN-classifier 
        while the darker-colored curves correspond to the embedding results.
        The success probability per gate $P_{\text{srpg}}$ is calculated using the equation \cref{eq:success_probability},
        while the success probability of entire circuit $P_{\text{success}}$ can be calculated by
        $P_{\text{success}} = P_{\text{srpg}}^{256-1}$, as we have $16\times16 - 1$ unitary gates in the circuit.
        All the metrics are for the training dataset.
        We also plot the schedule of $w$ with orange curve.
    }\label{fig:mnist_ftn}
\end{figure}

The results are shown in \cref{fig:mnist_ftn}.
The lighter-colored curves correspond to training the FTN-classifier 
with $d=10$ and $\chi=2$. 
These curves show that the FTN-classifier is trained effectively. 
The darker-colored curves indicate how the model evolves after the pre-trained 
FTN-classifier is embedded into a quantum circuit, following the 
adiabatic encoding procedure. 
Whenever $w$ is increased, the loss function experiences a discontinuous jump 
(because the embedding changes), 
but subsequent unitary gate optimization reduces the loss back to its 
previous level, demonstrating the effectiveness of our approach.
Additionally, the slight gap at the connection point between the light and dark lines 
arises from the initial jump in $w$, 
where $w$ moves from $0$ to $\Delta w$ immediately after embedding.
This causes a temporary increase in the loss, which is subsequently minimized.
The loss value and the accuracy for the FTN-classifier (before adiabatic encoding) and the qFTN-classifier (after adiabatic encoding) are shown in \cref{tab:mnist_ftn}.
You can observe that the qFTN-classifier achieves better loss and accuracy than the FTN-classifier.
We have to be cautious that it is not necessarily the case that the qFTN-classifier is more expressive 
than the original FTN-classifier; one might argue that it has simply undergone additional training. 
An important note to stress here, however, is that performance after adiabatic encoding has at least not degraded.
This indicates that the transition to a postselection-free quantum circuit has succeeded without sacrificing accuracy.

\begin{table*}[htbp]
    \centering
    \begin{tabular}{|c||c|c|c|c|}
        \hline
        Model & Train Loss & Train Accuracy & Test Loss & Test Accuracy \\
        \hline
        qFTN-classifier & 0.03516 & 99.07\% & 0.08877 & 97.35\% \\
        FTN-classifier & 0.06875 & 97.92\% & 0.12842 & 96.65\% \\
        \hline
    \end{tabular}
    \vspace{1em}
    \caption{\normalfont The loss and accuracy for the MNIST dataset.}
    \label{tab:mnist_ftn}
\end{table*}

\subsection{Results on CIFAR-10}
\begin{figure}[htp]
    \centering
    \includegraphics[width=0.5\textwidth]{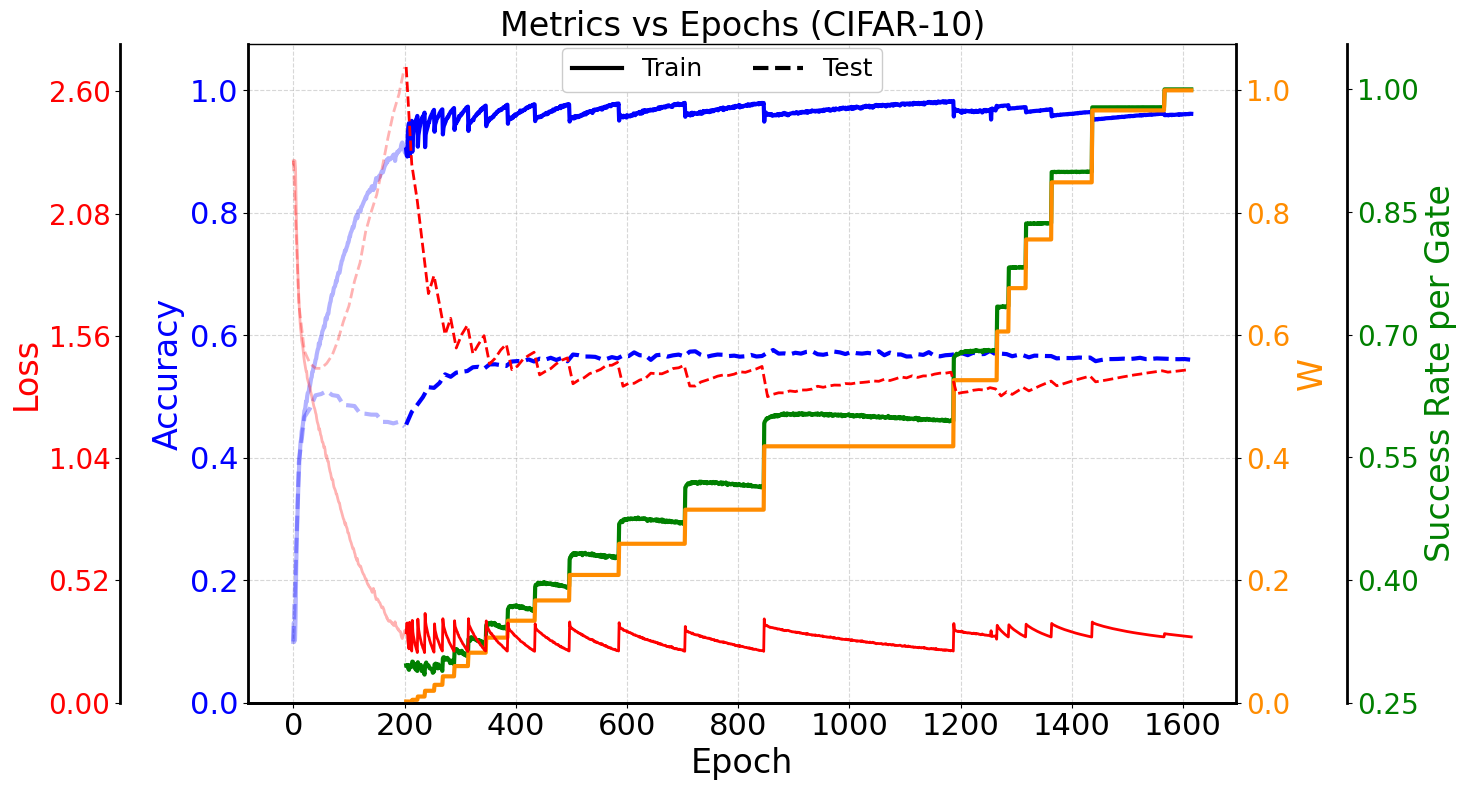}
    \caption{
        Training and embedding results for the CIFAR-10 dataset. 
        The color represents the same meaning as in \cref{fig:mnist_ftn} but this time we add the test results,
        which are displayed as dashed lines.
    }\label{fig:cifar_ftn}
\end{figure}

Next, we present our results on the CIFAR-10 dataset, which consists of $32\times32$ RGB color images. 
We downscale each image to $8\times8$, so each pixel $\bm{x} = (x_r, x_g, x_b)$ contains three values.
To accommodate this, we embed each scaled pixel value into a three-qubit state via 
$\phi(x_r)\otimes\phi(x_g)\otimes\phi(x_b)$, 
where $\phi(\cdot)$ is the local feature map from \cref{eq:feature_map_local}. 
This leads to TTNs with bond dimension $\chi = 8$, 
and the final linear layer becomes
$W: \mathbb{R}^{80} \to \mathbb{R}^{10}$.
The rest of the training and embedding procedures follow the same steps as in the MNIST case.
The results are shown in \cref{fig:cifar_ftn}.
We observe that the training accuracy for the CIFAR-10 dataset 
is reasonably high, whereas the corresponding test accuracy is much lower,
which indicates a clear overfitting. 
In this work, we primarily aimed to demonstrate that our method provides high expressive power while being trainable on a challenging task; therefore, 
we did not employ additional measures such as regularization, data augmentation, or early stopping, to enhance generalization~\cite{Zhang2017,Luo2018,Shorten2019,Xia2023}. 
Incorporating these techniques is an important direction for future research, 
particularly for real-world quantum machine learning applications. 
Nevertheless, it is promising that the adiabatic encoding of the FTN-classifiers retains or improves training performance.
It should also be noted that the qFTN-classifier outperforms the FTN-classifier on the test dataset, even though the training accuracy is improved.
Interestingly, while the training loss goes up at every update of $w$, the test loss goes down.
This anti-symmetric behavior helps the qFTN-classifier to achieve better test performance after the adiabatic encoding.
This potentially provides a numerical instance, although only a single training example, 
that supports the claim that QNNs can possess stronger generalization capabilities than purely classical models~\cite{Abbas2021,Caro2022}. 
In~\cref{tab:cifar_ftn}, we summarize the loss and accuracy of the models for the CIFAR-10 dataset.
\begin{figure}[htp]
    \centering
    \includegraphics[width=0.5\textwidth]{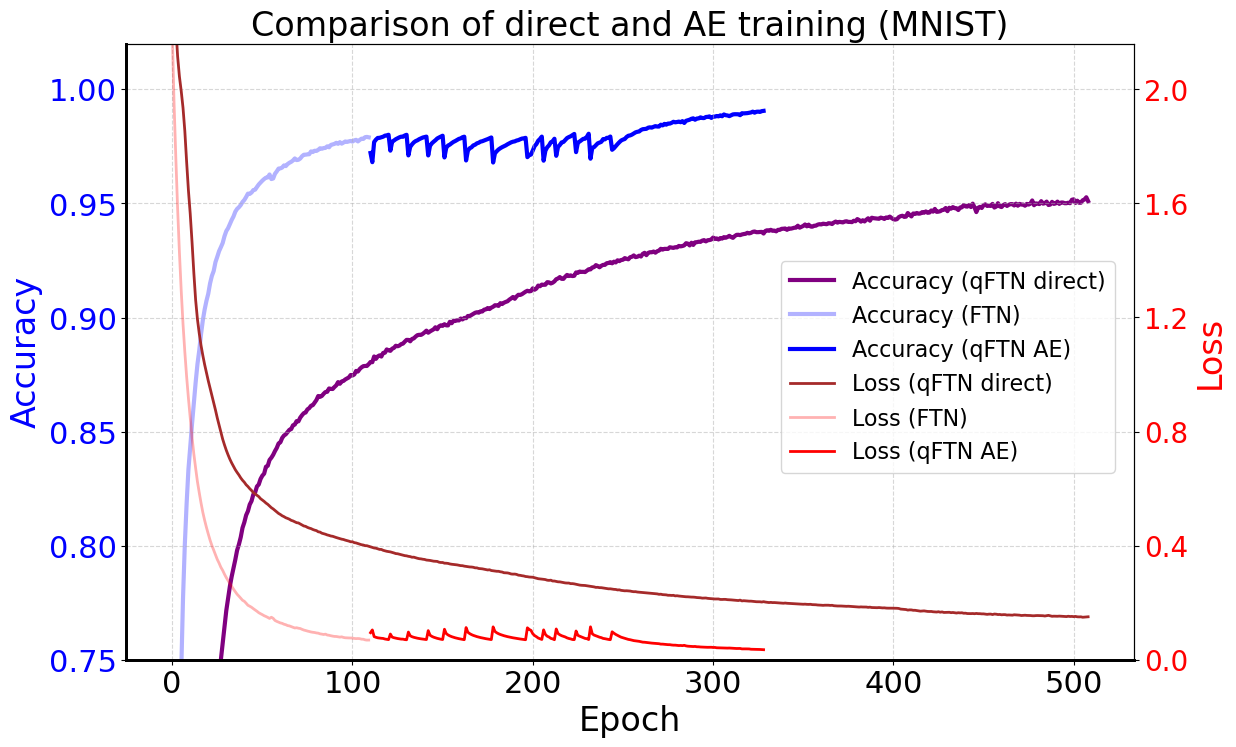}
    \caption{
        Comparison between the direct training of the qFTN-classifier and the adiabatic encoding (AE) for the MNIST dataset.
        The purple curve and brown curve corresponds to the accuracy and loss of the direct training of the qFTN-classifier respectively.
        The blue curve and red curve are the same as the the curves in \cref{fig:mnist_ftn}.
    }\label{fig:mnist_ftn_hr}
\end{figure}

\begin{table*}
    \centering
    \begin{tabular}{|c||c|c|c|c|}
        \hline
        Model & Train Loss & Train Accuracy & Test Loss & Test Accuracy \\
        \hline
        qFTN-classifier & 0.27996 & 96.16\% & 1.4176 & 56.11\% \\
        FTN-classifier & 0.29764 & 90.31\% & 2.7036 & 45.41\% \\
        \hline
    \end{tabular}
    \vspace{1em}
    \caption{\normalfont The loss and accuracy for the CIFAR-10 dataset.}
    \label{tab:cifar_ftn}
\end{table*}

\begin{figure*}[htp]
    \centering
    \includegraphics[width=0.9\textwidth]{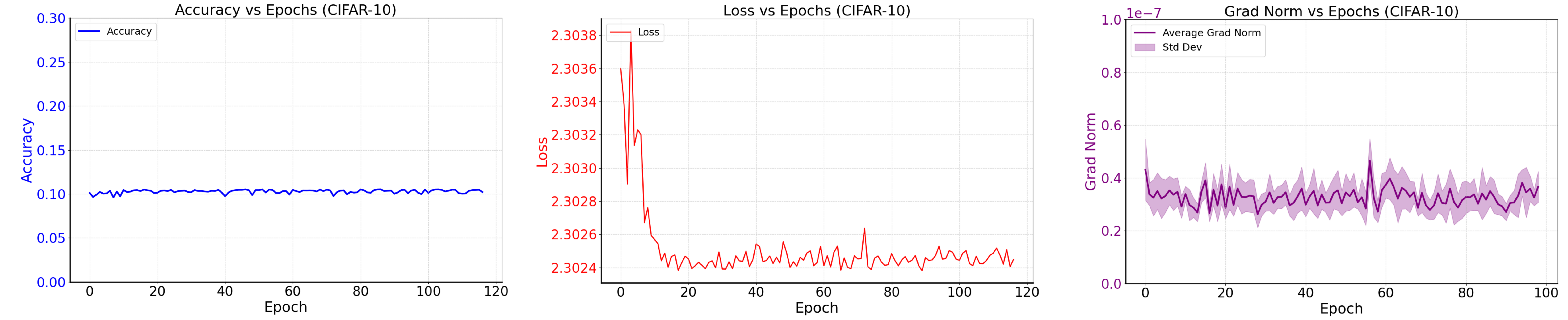}
    \caption{
        Comparison between the direct training of the qFTN-classifier and the adiabatic encoding for the CIFAR-10 dataset.
        The right figure is the norm of the gradient with respect to the first unitary matrix in the qTTN-classifiers (The unitary matrix directly attached to the first qubit). 
        The bold line is the average of the norm of the gradient over 10 qTTNs in qFTN, while
        the transparent area is the standard deviation over 10 samples.
    }\label{fig:cifar_ftn_hr}
\end{figure*}

\subsection{Directly Training qFTN-classifiers with Haar-random Initialization}

Our approach begins by training purely classical FTN-classifiers, 
embedding them into quantum circuits with postselection, 
and then removing postselection via adiabatic encoding.
This provides a sophisticated training strategy for qFTN-classifiers. 
However, a natural question arises: 
what if we initialize the qFTN-classifiers with Haar-random unitary gates 
and directly optimize them using Riemannian optimization from scratch? 
We address this question in this subsection.
The results for the MNIST dataset are shown in \cref{fig:mnist_ftn_hr}.
From these plots, we can observe two major drawbacks to directly training the qFTN-classifier, 
as opposed to starting from a trained FTN-classifier and applying adiabatic encoding:

\begin{enumerate}
    \item \textbf{Longer training time.} 
    Compared to training the FTN-classifier, direct training of the qFTN-classifier 
    requires about five times more epochs to converge. 
    Moreover, in terms of raw computational costs, 
    the classical TN-classifier scales as \(O(N \chi^3)\), 
    whereas the quantum circuit simulation scales as \(O(N \chi^6)\). 
    For \(\chi=2\), this can result in roughly 40 times more computational resources 
    for direct quantum-circuit-based training. 
    Even when eventual quantum hardware implementation is taken into account, these results suggest that adiabatic encoding offers a more efficient and less resource-intensive path to optimized qFTN-classifiers.
    
    \item \textbf{Local minima.}
    As seen in the figure, even though the directly trained qFTN-classifier and the adiabatically encoded qFTN-classifier theoretically share the same 
    expressive power, the final performance of the directly trained model is slightly worse. 
    This implies that direct training, at least in the present case, becomes trapped in local minima~\cite{You2021,Anschuetz2022}, 
    whereas adiabatic encoding leverages the pre-trained FTN-classifier initialization 
    to avoid such traps.
\end{enumerate}

Next, we present the results for the CIFAR-10 dataset,
when directly training the qFTN-classifier with Haar-random initializations and same training settings.
The results are shown in \cref{fig:cifar_ftn_hr}.
Clearly, the optimization encounters a \emph{barren plateau} problem, 
where gradients vanish and training is severely hindered
~\cite{McClean2018, Cerezo2021, Wang2021, Yao2025}.
Thus, our third drawback in directly training qFTN-classifiers is:

\begin{enumerate}
\setcounter{enumi}{2}
\item \textbf{Barren Plateaus.} 
While such plateaus did not appear in the previous dataset, 
they exist in CIFAR-10 training and almost completely halt the learning process. 
In principle, one expects that qTTN-circuits,
should suffer less from barren plateaus compared to other TTN-based QNN architectures.
However, in this scenario, we are optimizing many-qubit gates 
($\chi=8$ corresponds to effectively six-qubit gates), 
which must be decomposed into a large number of two-qubit gates. 
Hence, even if the circuit depth scales as $\log N$, 
the proportionality constant can become large enough to reintroduce 
barren plateau effects. 
Intuitively, 
barren plateaus become severe exponentially with the number $L$ of gates 
between the input and output. 
For qMPS, $L$ grows as $O(N)$ with the number of qubits $N$, 
whereas for qTTN, $L$ scales only as $O(\log N)$
~\cite{keisuke2025, Liu2022}. 
Nevertheless, once each many-qubit gate is decomposed into multiple two-qubit gates, 
the coefficient for $\log N$ becomes negligibly large, 
so that barren plateaus can still emerge in practice.
\end{enumerate}

\section{Discussion}\label{sec:discussion}
\subsection{Contributions}

In this work, we begin by introducing the forest tensor network classifiers, 
a multiclass classification framework consisting of multiple TTN-classifiers 
that collectively form a rich intermediate representation. 
This design ensures high expressive power while aiming for 
the eventual integration into quantum neural networks. 
These FTN-classifiers can be exactly represented by quantum neural networks
with postselection, but the success rate is exponentially small.
To circumvent this bottleneck, we employed adiabatic encoding
to systematically remove the exponential overhead of postselection
and encode to qFTN-classifiers that perform as well as the pre-trained FTN-classifiers,
without requiring any postselection. 
As a result, the final qFTN-classifiers are amenable to direct implementation 
on real quantum hardware.

To assess the effectiveness of our training strategy, 
we performed numerical simulations on two standard multiclass image classification tasks. 
In both cases, our approach trained successfully and achieved high training accuracy, thereby demonstrating that qFTN-classifiers not only possess strong expressive capability but are also readily trainable in practice.

\subsection{Future Work}
Future studies can provide deeper insights into the generalization behavior of our method.
Although our experiments demonstrated promising training performance 
on both datasets, there remains a gap between training and test results, 
particularly for CIFAR-10. Future work could investigate techniques 
such as regularization, data augmentation, or more advanced optimization 
methods to further improve test loss and accuracy.
Additionally, one could slightly modify our QNN architectures 
to incorporate translational symmetry and spatial locality, 
similar to the design principles of quantum convolutional networks~\cite{cong2019quantum}.
This may further improve the generalization performance of our approach.
Another direction is to extend our framework to use even more expressive TN architectures, 
such as MERA or PEPS.
While these architectures may be more complex to implement and train, 
they can offer broader applicability of quantum-classical hybrid machine learning.
Verifying the performance of our approach on physical quantum hardware 
is also a crucial next step. 
In this study, we embedded FTN-classifiers 
into qFTN-classifiers, classical-quantum hybrid QNNs.
Ultimately, however, our ambition is to use these QNNs merely as an initialization and then append further quantum gates to reach a regime in which the dynamics can no longer 
be efficiently simulated classically and thus achieve quantum advantage.
We regard this as an open challenge for future exploration.


\section*{Acknowledgments}
This work was supported by the Center of Innovation for Sustainable Quantum AI, JST Grant Number JPMJPF2221.
KM acknowledges the support of JSPS KAKENHI Grant Number JP24KJ0892.
TK acknowledges the support of JST SPRING Grant Number JPMJSP2108.

\bibliographystyle{IEEEtran}
\bibliography{main}  

\begin{thebibliography}{10}
\providecommand{\url}[1]{#1}
\csname url@samestyle\endcsname
\providecommand{\newblock}{\relax}
\providecommand{\bibinfo}[2]{#2}
\providecommand{\BIBentrySTDinterwordspacing}{\spaceskip=0pt\relax}
\providecommand{\BIBentryALTinterwordstretchfactor}{4}
\providecommand{\BIBentryALTinterwordspacing}{\spaceskip=\fontdimen2\font plus
\BIBentryALTinterwordstretchfactor\fontdimen3\font minus
  \fontdimen4\font\relax}
\providecommand{\BIBforeignlanguage}[2]{{%
\expandafter\ifx\csname l@#1\endcsname\relax
\typeout{** WARNING: IEEEtran.bst: No hyphenation pattern has been}%
\typeout{** loaded for the language `#1'. Using the pattern for}%
\typeout{** the default language instead.}%
\else
\language=\csname l@#1\endcsname
\fi
#2}}
\providecommand{\BIBdecl}{\relax}
\BIBdecl

\bibitem{White1992}
S.~R. White, ``Density matrix formulation for quantum renormalization groups,''
  \emph{Phys. Rev. Lett.}, vol.~69, no.~19, pp. 2863--2866, 1992.

\bibitem{Verstraete2008}
F.~Verstraete, V.~Murg, and J.~I. Cirac, ``Matrix product states, projected
  entangled pair states, and variational renormalization group methods for
  quantum spin systems,'' \emph{Advances in Physics}, vol.~57, no.~2, pp.
  143--224, 2008.

\bibitem{Orus2014}
R.~Orús, ``A practical introduction to tensor networks: Matrix product states
  and projected entangled pair states,'' \emph{Annals of Physics}, vol. 349,
  pp. 117--158, 2014.

\bibitem{Orus2019}
R.~Or{\'u}s, ``Tensor networks for complex quantum systems,'' \emph{Nature
  Reviews Physics}, vol.~1, no.~9, pp. 538--550, Sep 2019.

\bibitem{Cirac2021}
J.~I. Cirac, D.~P{\'e}rez-Garc{\'i}a, N.~Schuch, and F.~Verstraete, ``Matrix
  product states and projected entangled pair states: Concepts, symmetries,
  theorems,'' \emph{Rev. Mod. Phys.}, vol.~93, no.~4, p. 045003, 2021.

\bibitem{Oseledets2011}
I.~V. Oseledets, ``Tensor-train decomposition,'' \emph{SIAM Journal on
  Scientific Computing}, vol.~33, no.~5, pp. 2295--2317, 2011.

\bibitem{khoromskij2012}
B.~N. Khoromskij, ``Tensors-structured numerical methods in scientific
  computing: Survey on recent advances,'' \emph{Chemometrics and Intelligent
  Laboratory Systems}, vol. 110, no.~1, pp. 1--19, 2012.

\bibitem{cichocki2014}
A.~Cichocki, ``Tensor networks for big data analytics and large-scale
  optimization problems,'' \emph{arXiv preprint}, 2014.

\bibitem{nunez2022}
Y.~Núñez~Fernández, M.~Jeannin, P.~T. Dumitrescu, T.~Kloss, J.~Kaye,
  O.~Parcollet, and X.~Waintal, ``Learning feynman diagrams with tensor
  trains,'' \emph{Phys. Rev. X}, vol.~12, no.~4, p. 041018, 2022.

\bibitem{chan2002}
G.~K.-L. Chan and M.~Head-Gordon, ``Highly correlated calculations with a
  polynomial cost algorithm: A study of the density matrix renormalization
  group,'' \emph{J. Chem. Phys.}, vol. 116, no.~10, pp. 4462--4476, 2002.

\bibitem{chan2011}
G.~K.-L. Chan and S.~Sharma, ``The density matrix renormalization group in
  quantum chemistry,'' \emph{Annual Review of Physical Chemistry}, vol.~62, pp.
  465--481, 2011.

\bibitem{baiardi2020}
A.~Baiardi and M.~Reiher, ``The density matrix renormalization group in
  chemistry and molecular physics: Recent developments and new challenges,''
  \emph{J. Chem. Phys.}, vol. 152, no.~4, p. 040903, 2020.

\bibitem{nakatani2018}
N.~Nakatani, ``Matrix product states and density matrix renormalization group
  algorithm,'' in \emph{Reference Module in Chemistry, Molecular Sciences and
  Chemical Engineering}.\hskip 1em plus 0.5em minus 0.4em\relax Elsevier, 2018.

\bibitem{Pan2022}
F.~Pan and P.~Zhang, ``Simulation of quantum circuits using the big-batch
  tensor network method,'' \emph{Phys. Rev. Lett.}, vol. 128, p. 030501, Jan
  2022.

\bibitem{Liu2021}
Y.~A. Liu, X.~L. Liu, F.~N. Li, H.~Fu, Y.~Yang, J.~Song, P.~Zhao, Z.~Wang,
  D.~Peng, H.~Chen, C.~Guo, H.~Huang, W.~Wu, and D.~Chen, ``Closing the
  "quantum supremacy" gap: achieving real-time simulation of a random quantum
  circuit using a new sunway supercomputer,'' in \emph{Proceedings of the
  International Conference for High Performance Computing, Networking, Storage
  and Analysis}, ser. SC '21.\hskip 1em plus 0.5em minus 0.4em\relax New York,
  NY, USA: Association for Computing Machinery, 2021.

\bibitem{Patra2024}
S.~Patra, S.~S. Jahromi, S.~Singh, and R.~Or\'us, ``Efficient tensor network
  simulation of ibm's largest quantum processors,'' \emph{Phys. Rev. Res.},
  vol.~6, p. 013326, Mar 2024.

\bibitem{Fu2024}
R.~Fu, Z.~Su, H.-S. Zhong, X.~Zhao, J.~Zhang, F.~Pan, P.~Zhang, X.~Zhao, M.-C.
  Chen, C.-Y. Lu, J.-W. Pan, Z.~Pei, X.~Zhang, and W.~Ouyang, ``Surpassing
  sycamore: Achieving energetic superiority through system-level circuit
  simulation,'' in \emph{SC24: International Conference for High Performance
  Computing, Networking, Storage and Analysis}, 2024, pp. 1--20.

\bibitem{Vidal2003}
G.~Vidal, ``Efficient classical simulation of slightly entangled quantum
  computations,'' \emph{Phys. Rev. Lett.}, vol.~91, no.~14, p. 147902, 2003.

\bibitem{Han2018}
Z.-Y. Han, J.~Wang, H.~Fan, L.~Wang, and P.~Zhang, ``Unsupervised generative
  modeling using matrix product states,'' \emph{Phys. Rev. X}, vol.~8, p.
  031012, 2018.

\bibitem{Cheng2019}
S.~Cheng, L.~Wang, T.~Xiang, and P.~Zhang, ``Tree tensor networks for
  generative modeling,'' \emph{Phys. Rev. B}, vol.~99, p. 155131, Apr 2019.

\bibitem{Miller2021}
J.~Miller, G.~Rabusseau, and J.~Terilla, ``Tensor networks for probabilistic
  sequence modeling,'' \emph{Proceedings of the 24th International Conference
  on Artificial Intelligence and Statistics}, vol. 130, pp. 3079--3087, 2021,
  arXiv:2003.01039.

\bibitem{Stoudenmire2016}
E.~Stoudenmire and D.~J. Schwab, ``Supervised learning with tensor networks,''
  in \emph{Advances in Neural Information Processing Systems}, D.~Lee,
  M.~Sugiyama, U.~Luxburg, I.~Guyon, and R.~Garnett, Eds., vol.~29.\hskip 1em
  plus 0.5em minus 0.4em\relax Curran Associates, Inc., 2016.

\bibitem{Evenbly2019}
G.~Evenbly, ``Number-state preserving tensor networks as classifiers for
  supervised learning,'' \emph{arXiv preprint}, May 2019, proposes tensor
  networks built from number-state preserving tensors for classification tasks.

\bibitem{Reyes2020}
J.~Reyes and E.~M. Stoudenmire, ``A multi-scale tensor network architecture for
  classification and regression,'' \emph{arXiv preprint}, Jan 2020, combines
  wavelet transformations with MPS for supervised learning.

\bibitem{Reyes_2021}
J.~A. Reyes and E.~M. Stoudenmire, ``Multi-scale tensor network architecture
  for machine learning,'' \emph{Machine Learning: Science and Technology},
  vol.~2, no.~3, p. 035036, jul 2021.

\bibitem{Rudolph2023}
M.~S. Rudolph, J.~Miller, D.~Motlagh, J.~Chen, A.~Acharya, and
  A.~Perdomo-Ortiz, ``Synergistic pretraining of parametrized quantum circuits
  via tensor networks,'' \emph{Nature Communications}, vol.~14, no. 8367, Dec
  2023.

\bibitem{Iaconis2024}
J.~Iaconis, S.~Johri, and E.~Y. Zhu, ``Quantum state preparation of normal
  distributions using matrix product states,'' \emph{npj Quantum Information},
  vol.~10, no.~15, Jan 2024.

\bibitem{Wall2021-generative}
M.~L. Wall, M.~R. Abernathy, and G.~Quiroz, ``Generative machine learning with
  tensor networks: Benchmarks on near-term quantum computers,'' \emph{Phys.
  Rev. Res.}, vol.~3, p. 023010, Apr 2021.

\bibitem{Rudolph2024}
M.~S. Rudolph, J.~Chen, J.~Miller, A.~Acharya, and A.~Perdomo-Ortiz,
  ``Decomposition of matrix product states into shallow quantum circuits,''
  \emph{Quantum Science and Technology}, vol.~9, no.~1, p. 015012, Jan 2024.

\bibitem{Ran2020}
S.-J. Ran, ``Encoding of matrix product states into quantum circuits of one-
  and two-qubit gates,'' \emph{Phys. Rev. A}, vol. 101, p. 032310, Mar 2020.

\bibitem{Malz2024}
D.~Malz, G.~Styliaris, Z.-Y. Wei, and J.~I. Cirac, ``Preparation of matrix
  product states with log-depth quantum circuits,'' \emph{Phys. Rev. Lett.},
  vol. 132, p. 040404, Jan 2024.

\bibitem{Manabe2024}
H.~Manabe and Y.~Sano, ``The state preparation of multivariate normal
  distributions using tree tensor network,'' \emph{arXiv preprint}, Dec 2024.

\bibitem{Sugawara2025}
S.~Sugawara, K.~Inomata, T.~Okubo, and S.~Todo, ``Embedding of tree tensor
  networks into shallow quantum circuits,'' \emph{arXiv 2501.18856}, Jan 2025.

\bibitem{Green2025}
J.~Green and J.~B. Wang, ``Quantum encoding of structured data with matrix
  product states,'' \emph{arXiv preprint}, Feb 2025.

\bibitem{Beer2020}
K.~Beer, D.~Bondarenko, T.~Farrelly, T.~J. Osborne, R.~Salzmann,
  D.~Scheiermann, and R.~Wolf, ``Training deep quantum neural networks,''
  \emph{Nature Communications}, vol.~11, no. 808, Feb 2020, open access.

\bibitem{Abbas2021}
A.~Abbas, D.~Sutter, C.~Zoufal, A.~Lucchi, A.~Figalli, and S.~Woerner, ``The
  power of quantum neural networks,'' \emph{Nature Computational Science},
  vol.~1, pp. 403--409, Jun 2021.

\bibitem{Kwak2021}
Y.~Kwak, W.~J. Yun, S.~Jung, and J.~Kim, ``Quantum neural networks: Concepts,
  applications, and challenges,'' \emph{arXiv preprint}, Aug 2021, covers
  quantum deep learning principles, achievements, challenges, and future
  directions.

\bibitem{Peruzzo2014}
A.~Peruzzo, J.~McClean, P.~Shadbolt, M.-H. Yung, X.-Q. Zhou, P.~J. Love,
  A.~Aspuru-Guzik, and J.~L. O’Brien, ``A variational eigenvalue solver on a
  photonic quantum processor,'' \emph{Nature Communications}, vol.~5, no. 4213,
  Jul 2014.

\bibitem{Perdomo-Ortiz2018}
A.~Perdomo-Ortiz, M.~Benedetti, J.~Realpe-G\'{o}mez, and R.~Biswas,
  ``Opportunities and challenges for quantum-assisted machine learning in
  near-term quantum computers,'' \emph{Quantum Science and Technology}, vol.~3,
  no.~3, p. 030502, Jun 2018.

\bibitem{Shi2006}
Y.-Y. Shi, L.-M. Duan, and G.~Vidal, ``Classical simulation of quantum
  many-body systems with a tree tensor network,'' \emph{Phys. Rev. A}, vol.~74,
  p. 022320, Aug 2006.

\bibitem{Chen2023}
H.~Chen and T.~Barthel, ``Machine learning with tree tensor networks, cp rank
  constraints, and tensor dropout,'' \emph{IEEE Transactions on Neural Networks
  and Learning Systems}, vol.~34, pp. 697--710, 2023.

\bibitem{Nie2025}
C.~Nie, J.~Chen, and Y.~Chen, ``Deep tree tensor networks for image
  recognition,'' \emph{arXiv preprint}, Feb 2025.

\bibitem{Liu2019}
D.~Liu, S.-J. Ran, P.~Wittek, C.~Peng, R.~B. Garc\'ia, G.~Su, and
  M.~Lewenstein, ``Machine learning by unitary tensor network of hierarchical
  tree structure,'' \emph{New Journal of Physics}, vol.~21, no.~7, p. 073059,
  Jul 2019, open access.

\bibitem{Hikihara2023}
T.~Hikihara, H.~Ueda, K.~Okunishi, K.~Harada, and T.~Nishino, ``Automatic
  structural optimization of tree tensor networks,'' \emph{Phys. Rev. Res.},
  vol.~5, p. 013031, Jan 2023.

\bibitem{Harada2025}
K.~Harada, T.~Okubo, and N.~Kawashima, ``Tensor tree learns hidden relational
  structures in data to construct generative models,'' \emph{Machine Learning:
  Science and Technology}, vol.~6, no.~2, p. 025002, apr 2025.

\bibitem{Vartiainen2004}
J.~J. Vartiainen, M.~M\"ott\"onen, and M.~M. Salomaa, ``Efficient decomposition
  of quantum gates,'' \emph{Phys. Rev. Lett.}, vol.~92, p. 177902, Apr 2004.

\bibitem{Krol2024}
A.~M. {Krol} and Z.~{Al-Ars}, ``{Beyond Quantum Shannon: Circuit Construction
  for General n-Qubit Gates Based on Block ZXZ-Decomposition},'' \emph{arXiv
  e-prints}, p. arXiv:2403.13692, Mar. 2024.

\bibitem{Nakanishi2024}
K.~M. Nakanishi, T.~Satoh, and S.~Todo, ``Decompositions of multiple
  controlled-$z$ gates on various qubit-coupling graphs,'' \emph{Phys. Rev. A},
  vol. 110, p. 012604, Jul 2024.

\bibitem{Cross2019}
A.~W. Cross, L.~S. Bishop, S.~Sheldon, P.~D. Nation, and J.~M. Gambetta,
  ``Validating quantum computers using randomized model circuits,'' \emph{Phys.
  Rev. A}, vol. 100, p. 032328, Sep 2019.

\bibitem{wall2021}
M.~L. Wall and G.~D'Aguanno, ``Tree-tensor-network classifiers for machine
  learning: From quantum inspired to quantum assisted,'' \emph{Phys. Rev. A},
  vol. 104, no.~4, p. 042408, 2021.

\bibitem{lazzarin2022}
M.~Lazzarin, D.~E. Galli, and E.~Prati, ``Multi-class quantum classifiers with
  tensor network circuits for quantum phase recognition,'' \emph{Physics
  Letters A}, vol. 434, p. 128056, 2022.

\bibitem{araz2022}
J.~Y. Araz and M.~Spannowsky, ``Classical versus quantum: Comparing
  tensor-network-based quantum circuits on large hadron collider data,''
  \emph{Phys. Rev. A}, vol. 106, no.~6, p. 062423, Dec 2022.

\bibitem{huggins2019}
W.~Huggins, P.~Patil, B.~Mitchell, K.~B. Whaley, and E.~M. Stoudenmire,
  ``Towards quantum machine learning with tensor networks,'' \emph{Quantum
  Science and Technology}, vol.~4, no.~2, p. 024001, 2019.

\bibitem{rieser2023}
H.-M. Rieser, F.~Köster, and A.~P. Raulf, ``Tensor networks for quantum
  machine learning,'' \emph{Proceedings of the Royal Society A: Mathematical,
  Physical and Engineering Sciences}, vol. 479, no. 2279, p. 20230218, 2023.

\bibitem{kodama2025}
N.~X. Kodama, A.~Bocharov, and M.~P. da~Silva, ``Image classification by
  combining quantum kernel learning and tensor networks,'' \emph{Phys. Rev. A},
  vol. 111, no.~1, p. 012630, 2025.

\bibitem{keisuke2025}
K.~Murota, ``Adiabatic encoding of pre-trained mps classifiers into quantum
  circuits,'' \emph{arXiv preprint arXiv:2504.09250}, 2025.

\bibitem{Vidal2008}
G.~Vidal, ``Class of quantum many-body states that can be efficiently
  simulated,'' \emph{Phys. Rev. Lett.}, vol. 101, p. 110501, Sep 2008.

\bibitem{Verstraete2004}
F.~{Verstraete} and J.~I. {Cirac}, ``{Renormalization algorithms for
  Quantum-Many Body Systems in two and higher dimensions},'' \emph{arXiv
  e-prints}, pp. cond--mat/0\,407\,066, Jul. 2004.

\bibitem{Iten2016}
R.~Iten, R.~Colbeck, I.~Kukuljan, J.~Home, and M.~Christandl, ``Quantum
  circuits for isometries,'' \emph{Phys. Rev. A}, vol.~93, p. 032318, Mar 2016.

\bibitem{Shende2004}
V.~V. Shende, I.~L. Markov, and S.~S. Bullock, ``Minimal universal two-qubit
  controlled-not-based circuits,'' \emph{Phys. Rev. A}, vol.~69, p. 062321, Jun
  2004.

\bibitem{Low2019hamiltonian}
G.~H. Low and I.~L. Chuang, ``Hamiltonian {S}imulation by {Q}ubitization,''
  \emph{{Quantum}}, vol.~3, p. 163, Jul. 2019.

\bibitem{chakraborty2019}
S.~Chakraborty, A.~Gily\'{e}n, and S.~Jeffery, ``{The Power of Block-Encoded
  Matrix Powers: Improved Regression Techniques via Faster Hamiltonian
  Simulation},'' in \emph{46th International Colloquium on Automata, Languages,
  and Programming (ICALP 2019)}, ser. Leibniz International Proceedings in
  Informatics (LIPIcs), C.~Baier, I.~Chatzigiannakis, P.~Flocchini, and
  S.~Leonardi, Eds., vol. 132.\hskip 1em plus 0.5em minus 0.4em\relax Dagstuhl,
  Germany: Schloss Dagstuhl -- Leibniz-Zentrum f{\"u}r Informatik, 2019, pp.
  33:1--33:14.

\bibitem{Holmes2023}
Z.~Holmes, N.~J. Coble, A.~T. Sornborger, and Y.~b.~u. Suba\ifmmode
  \mbox{\c{s}}\else \c{s}\fi{}\ifmmode \imath \else~\i \fi{}, ``Nonlinear
  transformations in quantum computation,'' \emph{Phys. Rev. Res.}, vol.~5, p.
  013105, Feb 2023.

\bibitem{Bonnabel2013}
S.~Bonnabel, ``Stochastic gradient descent on riemannian manifolds,''
  \emph{IEEE Transactions on Automatic Control}, vol.~58, no.~9, pp.
  2217--2229, Sep 2013.

\bibitem{geoopt2020kochurov}
M.~Kochurov, R.~Karimov, and S.~Kozlukov, ``Geoopt: Riemannian optimization in
  pytorch,'' 2020.

\bibitem{Zhang2017}
C.~Zhang, S.~Bengio, M.~Hardt, B.~Recht, and O.~Vinyals, ``Understanding deep
  learning requires rethinking generalization,'' \emph{International Conference
  on Learning Representations (ICLR)}, 2017, published in ICLR 2017.

\bibitem{Luo2018}
P.~Luo, X.~Wang, W.~Shao, and Z.~Peng, ``Towards understanding regularization
  in batch normalization,'' \emph{International Conference on Learning
  Representations (ICLR)}, 2018, arXiv:1809.00846 [cs.LG].

\bibitem{Shorten2019}
C.~Shorten and T.~M. Khoshgoftaar, ``A survey on image data augmentation for
  deep learning,'' \emph{Journal of Big Data}, vol.~6, no.~60, Jul 2019.

\bibitem{Xia2023}
Z.~Xia, ``Overfitting of cnn model in cifar-10: Problem and solutions,''
  \emph{Journal of Machine Learning Research}, vol.~37, p. 20230511, 2023, *
  Author to whom correspondence should be addressed.

\bibitem{Caro2022}
M.~Caro, H.-Y. Huang, M.~Cerezo, K.~Sharma, A.~T. Sornborger, P.~J. Coles, and
  L.~Cincio, ``Generalization in quantum machine learning from few training
  data,'' \emph{Nature Communications}, vol.~13, no.~1, p. 4919, 2022.

\bibitem{You2021}
X.~You and X.~Wu, ``Exponentially many local minima in quantum neural
  networks,'' in \emph{Proceedings of the 38th International Conference on
  Machine Learning (ICML)}, ser. Proceedings of Machine Learning Research, vol.
  139, 2021, pp. 12\,192--12\,202.

\bibitem{Anschuetz2022}
E.~R. Anschuetz and B.~T. Kiani, ``Quantum variational algorithms are swamped
  with traps,'' \emph{Nature Communications}, vol.~13, no.~1, p. 7760, 2022.

\bibitem{McClean2018}
J.~R. McClean, S.~Boixo, V.~N. Smelyanskiy, R.~Babbush, and H.~Neven, ``Barren
  plateaus in quantum neural network training landscapes,'' \emph{Nature
  Communications}, vol.~9, no.~1, p. 4812, 2018.

\bibitem{Cerezo2021}
M.~Cerezo, A.~Sone, T.~Volkoff, L.~Cincio, and P.~J. Coles, ``Cost function
  dependent barren plateaus in shallow parametrized quantum circuits,''
  \emph{Nature Communications}, vol.~12, no.~1, p. 1791, 2021.

\bibitem{Wang2021}
S.~Wang, E.~Fontana, M.~Cerezo \emph{et~al.}, ``Noise-induced barren plateaus
  in variational quantum algorithms,'' \emph{Nature Communications}, vol.~12,
  no.~1, p. 6961, 2021.

\bibitem{Yao2025}
Y.~Yao and Y.~Hasegawa, ``Direct gradient computation of parameterized quantum
  circuits,'' \emph{arXiv preprint}, Mar 2025.

\bibitem{Liu2022}
Z.~Liu, L.-W. Yu, L.-M. Duan, and D.-L. Deng, ``Presence and absence of barren
  plateaus in tensor-network based machine learning,'' \emph{Phys. Rev. Lett.},
  vol. 129, p. 270501, Dec 2022.

\bibitem{cong2019quantum}
I.~Cong, S.~Choi, and M.~D. Lukin, ``Quantum convolutional neural networks,''
  \emph{Nature Physics}, vol.~15, pp. 1273--1278, 2019.

\end{thebibliography}

\end{document}